\documentclass[11pt,english]{article}
\usepackage{bm,bbm,lscape}
\usepackage{amsmath,authblk}
\usepackage[authoryear,round]{natbib}
\usepackage{marvosym}
\usepackage{multicol}
\usepackage[british]{babel}
\usepackage{pstricks,graphicx,xcolor,geometry,setspace,lineno}
\usepackage[british]{babel}
\usepackage[justification=justified,singlelinecheck=false,labelfont=it]{caption}
\usepackage{fancyhdr}
\title{Modelling group dynamic animal movement}
\author[1]{\normalsize Roland Langrock\footnote{Corresponding author. E-mail: \texttt{roland@mcs.st-and.ac.uk}. Address: CREEM, The Observatory, Buchanan Gardens, University of St Andrews, St Andrews, KY16 9LZ, UK.}}
\author[1,2]{J.\ Grant C.\ Hopcraft}
\author[3]{Paul G.\ Blackwell}
\author[4,5]{Victoria Goodall}
\author[1]{Ruth King}
\author[3]{Mu Niu}
\author[6]{Toby A.\ Patterson}
\author[7]{Martin W.\ Pedersen}
\author[8]{Anna Skarin}
\author[1]{Robert S.\ Schick}

\affil[1]{\tiny Center for Research into Ecological and Environmental Modelling, School of Mathematics and Statistics, University of St Andrews, UK.\normalsize\vspace{-0em}}
\affil[2]{\tiny Boyd Orr Centre for Population and Ecosystem Health, College of Medical Veterinary and Life Sciences, University of Glasgow, UK.\normalsize\vspace{-0em}}
\affil[3]{\tiny School of Mathematics and Statistics, University of Sheffield, UK.\normalsize\vspace{-0em}}
\affil[4]{\tiny South African Environmental Observation Network, Fynbos Node, South Africa.\normalsize\vspace{-0em}} 
\affil[5]{\tiny School of Statistics and Actuarial Science, University of the Witwatersrand, South Africa.\normalsize\vspace{-0em}}
\affil[6]{\tiny Commonwealth Scientific and Industrial Research Organisation, Wealth from Oceans research Flagship, Hobart, Australia. \normalsize\vspace{-0em}}
\affil[7]{\tiny National Institute of Aquatic Resources, Technical University of Denmark, Charlottenlund, Denmark.\normalsize\vspace{-0em}}
\affil[8]{\tiny Department of Animal Nutrition and Management, Swedish University of Agricultural Sciences, Uppsala, Sweden.\normalsize}

\date{}

\begin{document}
 
\begin{spacing}{1}
\maketitle

 \begin{linenumbers}
 \setlength\linenumbersep{1cm}
 \rightlinenumbers
\vspace{-1em}

\hrule
 \vspace{2em}
 
\noindent
{\bf Summary}\\
\noindent
{\bf 1.} Group dynamic movement is a fundamental aspect of many species' movements. The need to adequately model individuals' interactions with other group members has been recognised, particularly in order to differentiate the role of social forces in individual movement from environmental factors. However, to date, practical statistical methods which can include group dynamics in animal movement models have been lacking.\\
\noindent
{\bf 2.} We consider a flexible modelling framework that distinguishes a group-level model, describing the movement of the group's centre, and an individual-level model, such that each individual makes its movement decisions relative to the group centroid. The basic idea is framed within the flexible class of hidden Markov models, extending previous work on modelling animal movement by means of multi-state random walks.\\
\noindent
{\bf 3.} While in simulation experiments parameter estimators exhibit some bias in non-ideal scenarios, we show that generally the estimation of models of this type is both feasible and ecologically informative.\\ 
\noindent
{\bf 4.} We illustrate the approach using real movement data from 11 reindeer {\it (Rangifer tarandus)}. Results indicate a directional bias towards a group centroid for reindeer in an encamped  state. Though the attraction to the group centroid is relatively weak, our model successfully captures group-influenced movement dynamics. Specifically, as compared to a regular mixture of correlated random walks, the group dynamic model more accurately predicts the non-diffusive behaviour of a cohesive mobile group. \\
\noindent
{\bf 5.} As technology continues to develop it will become easier and less expensive to tag multiple individuals within a group in order to follow their movements. Our work provides a first inferential framework for understanding the relative influences of individual versus group-level movement decisions. This framework can be extended to include covariates corresponding to environmental influences or body condition. As such, this framework allows for a broader understanding of the many internal and external factors that can influence an individual's movement.

\vspace{1em}
 \vspace{1em}

\section*{Introduction}\label{intro}

In ecology, there is growing interest in understanding and modelling animal movement \citep{nat08}. For example, understanding movement patterns is critical when considering the potential consequences of land use, climate change or anthropogenic activities on range expansions, the spread of invasive species, or movement of hosts and their pathogens into new vulnerable areas \citep{bow05}. In addition, quantitative descriptions of animals' movements may also contribute to our understanding of the underlying movement decisions (e.g., \citealp{sch08}). Many of these decisions are driven by social factors \citep{eft07,hay08}, which highlights the importance of being able to adequately model group dynamic movement patterns \citep{mor10}.

Following advances in tracking technology, recent years have seen a fast growing body of literature concerned with the statistical modelling of animal movement. Different modelling approaches have been extensively discussed, comprising, {\em inter alia}, state-space models \citep{jon05,pat08} and stochastic differential equations \citep{bla03,pre04}. Such approaches have mostly neglected potential interactions between different animals and, until now, most studies have assumed that the movement of one individual within a group is representative of the group's overall movement \citep{mor10}. However, animals often do not move independently of each other, and therefore analysing the movement of individual animals without considering the dynamics of the group could be misleading. %For example, if animals are considered to be non-interacting particles, there is the potential to falsely relate their patterns of movement to environmental conditions when in fact their interactions with other animals -- either conspecifics, prey or predators -- drive their movement and behaviour.
Reduced production costs and miniaturisation of tracking technology mean that field researchers can monitor the movement of many more individuals than was previously possible, including simultaneously tracking several individuals within the same group. These new advances provide the opportunity for studying the inter- and intra-group dynamics within a population.

The movement of individuals in a group is typically investigated using self-propelled particle (SPP) models that capture the alignment between neighbours in self-organised swarms.  For SPP models to maintain coordinated group movement all individuals must adhere to basic mechanistic rules in which the forces of attraction (e.g., social interactions such as information sharing or vigilance) and repulsion (e.g., avoiding collisions with neighbours) are optimised within an interaction zone \citep{buh06,man11,str11}. In classic Lagrangian SPP models, individuals match their speed and alignment at discrete time intervals and are virtually homologous copies of one another, with the exception that some may be ``informed'' while others ``na{\"\i}ve'' \citep{cou05,con09}.  

While in the existing literature on SPP models the aim is to replicate movement patterns in forward simulations by defining certain rules of behaviour, we suggest a novel approach which fundamentally differs from SPP models in that our model is fitted statistically to telemetry data. Our modelling approach allows individuals to switch between different behavioural states, so that they can either be gregarious members of a group under certain conditions such as when they are exposed to risks or are uninformed, but then break away from the group once conditions change in order to capitalise on resources when competition is intense.  Rather than individuals being attracted to their neighbours as in models of collective movement \citep{str11}, our approach is to capture group fission-fusion dynamics where individuals are periodically attracted to an abstract point, which we call the {\em group centroid}. This centroid could be the centre of mass of the group (a proxy for social networks, as described by \citealp{cro10}) or a dominant ``informed'' alpha animal that leads the group (as described by \citealp{nag12}). The two approaches potentially overlap, in that it can be shown that some SPP-type models, for example certain models with symmetric pairwise interaction between individuals, are equivalent to particular cases of our centroid-based models. This gives further motivation for our approach, but we do not pursue the cases of equivalence in detail here.
  
Here we formulate the approach within an easily accessible framework, given by the class of hidden Markov models (HMMs) \citep{zuc09}. We show the feasibility of the HMM-based approach by means of simulation before fitting the model to real data on the movement of 11 reindeer {\it (Rangifer tarandus)}.

\section*{Materials and methods}

\subsection*{\small {\it GENERAL FORMULATION OF THE MODELLING APPROACH}}

We consider a modelling framework with a parent-child structure. First, a group-level model describes movement of some entity  (the ``parent'') that characterises the group's centre of gravity and drives the movement behaviour of individuals. Depending on the system this entity can have different interpretations: e.g., it may be the group centre (represented by the mathematical centroid), or the location of the group leader. This entity is a tool that allows for modelling correlation between the individuals' movement paths in a flexible yet intuitive way. It is crucial that the entity adequately represents the point relative to which individual animals make their movement decisions. Hereafter we will refer to this entity as the {\em (group) centroid}. Second, at the individual level, the animals (the ``children'') make their movement decisions relative to the group centroid. Such decisions can involve attraction to the centroid, repulsion from the centroid, or disregard of the centroid. %More detailed descriptions of models for both the group centroid's and the individual-level movement are given below. 

The suggested concept is immensely flexible and, in principle, can be implemented by means of different stochastic models. Here we use a discrete-time HMM-based approach for observations that are regularly spaced in time. In these instances it is convenient to model the bivariate time series of step lengths (between successive locations) and turning angles (between successive movement directions); see \citet{mor04}. 
 
\subsection*{\small {\it THE BUILDING BLOCKS -- CORRELATED AND BIASED RANDOM WALKS}}

Our model for group movement is composed of well-known movement models: correlated and biased random walks (CRWs and BRWs, respectively), and walks that are both correlated \textit{and} biased (BCRWs) \citep{cod08,lan12}. CRWs involve positive (or negative) correlation in direction. In discrete time, they can be expressed by a turning angle distribution with mass centred around zero (or $\pi$). Biasedness of random walks can either refer to a general preference for some direction (e.g., East) or a bias towards a particular location. For example, in discrete time, a bias towards the location $(x^{(c)},y^{(c)})$ is obtained by assuming that the expected movement direction at time $t+1$ is the direction of the vector $(x^{(c)},y^{(c)})-(x_t,y_t)$, where $(x_t,y_t)$ is the animal's location at time $t$. In our model, the location a given animal is attracted to will vary over time, denoted by $(x_t^{(c)},y_t^{(c)})$, $t=1,\ldots,T$, so that $(x_t^{(c)},y_t^{(c)})$ is the location of the centroid at time $t$. An animal may be attracted to the centroid (in which case the expected movement direction at time $t+1$ is the direction of the vector $(x_{t+1}^{(c)},y_{t+1}^{(c)})-(x_t,y_t)$), or it may be repulsed by the centroid (in which case the expected movement direction is the direction of the vector $(x_t,y_t)-(x_{t+1}^{(c)},y_{t+1}^{(c)})$), or it may move in disregard of the centroid (e.g., it may move according to a CRW, or according to a BRW with a fixed centre of attraction).  

\subsection*{\small {\it INDIVIDUAL-LEVEL AND CENTROID MOVEMENT MODELS}}

While single-state random walk models can be appropriate to describe animal movement on short temporal scales, at longer time scales they are usually too inflexible because the animal changes its movement pattern according to changes of its behavioural state \citep{mor04}. In contrast, in multi-state random walks -- which are HMMs -- the movement pattern of the animal is assumed to depend on the current underlying behavioural state of the animal (e.g., foraging, migrating, resting). The state sequence in these models is generated by a Markov chain, which leads to temporal autocorrelation in both the behavioural states and in the observed movement patterns. Notationally, we summarize the state transition probabilities of the (homogeneous) Markov chain in the $N \times N$ matrix $\boldsymbol{\Gamma}=\bigl( \gamma_{ij}  \bigr)$, where $\gamma_{ij}$ denotes the probability of the animal switching from state $i$ (at any time $t$) to state $j$ (at time $t+1$). 

In a multi-state random walk, say with $N$ states, each of the $N$ states is associated with a distinct random walk pattern (CRW, BRW or BCRW). For more details and discussion on multi-state random walks in the context of movement modelling, we refer the reader to \citet{pat09} and \citet{lan12}. In this manuscript, we focus on (individual-level) multi-state random walks in which at least one of the $N$ states involves either a BRW or a BCRW with bias relative to the centroid's location (either positive or negative). Crucially, if the movement of multiple individuals is considered, and the movement models all involve a BRW or BCRW relating to the same group centroid, then the collection of all individual-level models can capture various degrees of possible correlation of the multiple movement paths. For example, individuals may differ in their bias towards the centroid, but as long as they exhibit some tendency towards the centroid, the paths will be correlated.

In order to simulate and predict from such individual-level models, one also needs a model for the movement of the group centroid. The bivariate time series corresponding to the centroid's movement can, in principle, also be modelled using (dependent) mixtures of random walks, i.e., HMMs. However, the location of the centroid often cannot be directly observed, and we address this issue below.

\subsection*{\small {\it MODEL FITTING}}

The likelihood of an HMM can conveniently be calculated using an efficient recursive scheme called the {\em forward algorithm}, which effectively corresponds to a summation over all possible state sequences.
Thus, the maximum likelihood estimates (MLEs) of the parameters can usually easily be obtained by direct numerical likelihood maximization (for example using \texttt{nlm} in R). In the case where the parameters are constant over time, the likelihood is given by
\begin{linenomath*}
\begin{equation}\label{likfor}
\mathcal{L}= \boldsymbol{\delta}^{(1)}\mathbf{P}(\mathbf{z}_1)\boldsymbol{\Gamma}\mathbf{P}(\mathbf{z}_2)\boldsymbol{\Gamma}
\cdot \ldots \cdot \boldsymbol{\Gamma}\mathbf{P}(\mathbf{z}_{T-1})\boldsymbol{\Gamma}\mathbf{P}(\mathbf{z}_T)\mathbf{1} \, . \vspace{-0em}
\end{equation}
\end{linenomath*}
Here $\mathbf{P}(\mathbf{z})= \text{diag} \bigl(   f_1(\mathbf{z}) , \ldots , f_N(\mathbf{z}) \bigr)$, with $f_n$ denoting the state-dependent probability density function of the observations, given the animal is in state $n$, and $\mathbf{z}_1, \mathbf{z}_2, \ldots, \mathbf{z}_T$ is the bivariate series of observed step lengths and turning angles. The density $f_n$ is determined by the type of random walk assumed in state $n$ and the state-dependent distributions considered for step lengths and turning angles (cf.\ \citealp{lan12}). Furthermore, $\mathbf{1}$ is a column vector of ones and $\boldsymbol{\delta}^{(1)}$ denotes the initial distribution of the Markov chain.

If one uses multi-state random walks to model both centroid and individual-level movement, then expression (\ref{likfor}) applies to both the centroid and the individual-level movement models, in each case with $\mathbf{P}$, $\boldsymbol{\Gamma}$ and $\boldsymbol{\delta}^{(1)}$ defined according to the model formulation. However, in both the centroid and individual-level cases, in the evaluation of (\ref{likfor}) the locations of the group centroid at the observation times are required. The determination of these locations depends on the type of centroid considered, and may be straightforward in some cases but intractable in others. For example, if the centroid corresponds to an animal leading the entire group, then it may be possible to identify and tag that animal, thus yielding the centroid's locations directly. 
 
We focus here on the case where the centroid roughly corresponds to the core of the group, i.e., a virtual entity essentially corresponding to the mathematical centroid of the group. If multiple individuals within the same group are tagged, then at any given time $t$ we simply compute the mean of all observed individuals' locations at that time, and consider this mean as an estimate of the centroid's location. If the group of animals is such that individuals occasionally separate from the core group, yet are still nominally associated with the group, then we expect that using an (outlier-)robust mean of the locations leads to a more accurate approximation of such a centroid's location. To do this, we used the minimum volume ellipsoid estimator provided in the function \texttt{cov.rob} from the R-package \texttt{MASS}. % \citep{rou90}.  

\subsection*{\small {\it EXPERIMENTS WITH SIMULATED DATA}}
In order to evaluate the performance of the proposed model, we initially conduct a simulation study. We simulate data from the proposed group dynamic movement model in an HMM framework, generating a movement path for the centroid comprising $T=250$ locations, and, based on the simulated centroid locations, $20$ individual movement paths, again each comprising $T=250$ locations. The movement model for the centroid is a BCRW, with some bias towards the location $(0,0)$, but mainly positive correlation in successive movement directions (for details see Table S1). Each individual-level movement path is generated by a two-state HMM, with state 1 a BRW (bias towards the moving centroid) and state 2 a CRW (positive correlation in successive movement directions). These states correspond to desire to stay within the group (state 1), and desire to forage independently of the group (state 2). We use gamma distributions to generate step lengths (letting $\mu_i$ and $\sigma_i$ denote the mean and standard deviation, respectively, in state $i$), and von Mises distributions to generate turning angles (letting $\nu_i$ and $\kappa_i$ denote the mean and concentration, respectively, in state $i$). We consider two different simulation scenarios, with the following parameter values considered for the individual-level movement models:
\begin{itemize}
\item[] {\bf Scenario A} \\ 
State 1: $\mu_1=30$, $\sigma_1=15$, $\nu_1$ determined by centroid location, $\kappa_1=2.5$  \\
State 2: $\mu_2=50$, $\sigma_2=25$, $\nu_2=0$ (not estimated), $\kappa_2=5$
\item[] {\bf Scenario B} \\
State 1: $\mu_1=30$, $\sigma_1=15$, $\nu_1$ determined by centroid location, $\kappa_1=2.5$  \\
State 2: $\mu_2=30$, $\sigma_2=15$, $\nu_2=0$ (not estimated), $\kappa_2=5$
\end{itemize}
In both scenarios,  
$$ \boldsymbol{\Gamma} = \begin{pmatrix} 0.95 & 0.05 \\ 0.15 & 0.85 \end{pmatrix} .$$
This state process is such that individuals spend most of the time following the centroid (in state 1, occupied $75\%$ of the time according to the stationary distribution), but occasionally split from the group and move solitarily (in state 2). Scenario A represents a setting in which the identification of the two hidden behavioural states based on the observations is expected to be accurate (since not only the turning angle distributions, but also the step length distributions differ across states), and Scenario B represents a setting in which the identification of the states is expected to be more challenging (since only the turning angle distributions differ across states, i.e., only the attraction to the centroid distinguishes state 1 from state 2 at the observation level). %In both scenarios we use the same BCRW to generate the centroid's locations.  

By numerically maximizing the respective likelihood, we fit the model to the centroid's series of step lengths and turning angles (initially assuming that these are observed), and then simultaneously to all $20$ individual-level series of step lengths and turning angles, assuming common parameters for all individuals.  We also explore how well the estimation works in cases where the true location of the centroid is unknown, and hence can only be approximated. To do this, we fit the centroid and the individual-level movement model using two different approximations of the centroid's locations: 1) the centroid calculated as the simple mean of all individuals' locations at each time; and 2) the centroid calculated as the robust mean of the individuals' locations at each time. 

This whole simulation and model fitting exercise was repeated 500 times, and performance is evaluated based on summary statistics of the parameter estimates. In each scenario, for each simulation run and three estimation methods (corresponding to different series of exact or approximate centroid locations), we use the Viterbi algorithm (see Chapter 5 in \citealp{zuc09}) to find the sequence of states $s_1^*,\ldots,s_T^*$ that under the fitted model is most likely to have given rise to the observed sequence, 
\begin{linenomath*}
\begin{equation*}
(s_1^*,\ldots, s_T^*)= \underset{(s_1,\ldots, s_T) \in \{1,2\}^T}{\operatorname{argmax}} \Pr (s_1,\ldots,s_T \, | \, z_1, \ldots, z_t), 
\end{equation*}
\end{linenomath*}
and establish the proportion of states correctly identified. %The R code used in the simulation experiments is provided in the supplementary material (S2), as is an example movie illustrating the movement pattern of the individuals and the centroid in Scenario B (S3).

\subsection*{\small {\it APPLICATION TO REAL DATA}}

We fit our model to location data of reindeer. The annual migration of reindeer follows a seasonal progression of snow-melt and fresh vegetative growth that broadly describes the general movement pattern of the population \citep{ska08,ska10}. However, herds often follow slightly different paths each year and individuals within these herds make independent decisions based on the environmental conditions they encounter {\it en route}. Although an individual reindeer may reduce its grazing competition by moving away from the herd, it also stands a greater chance of being killed by predators and therefore the choice an individual reindeer makes about how and where to move is balanced between finding enough food for itself but also by staying within the safety of the group.

We consider location data for 11 female reindeer, recorded in June and July 2003 (which is in the post-calving period) with GPS collars. For each of the 11 individuals, a maximum of 225 hourly fixes is considered (corresponding to $9.4$ days of observation), with very few missing data ($0.2\%$). From exploratory analysis, we deduce that these 11 individuals belong to the same herd. Although the total number of individuals in the herd varies over time, a core group tended to stay together for the period we considered. 

Since individual reindeer occasionally separate from the group, we use a robust mean of the individual's locations observed at a given time to approximate the centroid location at that time. Modelling the movement of the (approximated) centroid requires only standard methods, but fitting ecologically complex and hence informative models at the centroid level is difficult in this case as the corresponding time series is short. Although resources might change very rapidly, we do not have the spatial data required to capture these short-term fluxes. Therefore, we focus exclusively on the individual-level model in this analysis. For simplicity, and for the sake of parsimony in terms of the number of parameters, we assume the parameters of the individual-level model to be common to all individuals. We fit a two-state model, with state 1 involving a BRW with bias towards the (moving) centroid, and state 2 involving a CRW such that the turning angle distribution has mean zero, i.e., with the reindeer tending to move straight ahead. We use gamma distributions for modelling the step lengths, and von Mises distributions for modelling the turning angles. 

From the suggested class of HMMs for movement at the individual level, this is the simplest example of a group dynamic movement model that still allows for the individuals to occasionally separate from the group. The purpose of the analysis is to illustrate that the suggested type of movement models can successfully be applied to real data, and that this can lead to biologically interesting insights. As the model fitting is illustrative, we do not search for the optimal model from a suite of candidate models. We do however check the adequacy of the described model by computing forecast pseudo-residuals for both step lengths and turning angles, as described in \citealt{lan12}, separately for each of the 11 individuals, resulting in 22 series of residuals. Such residuals are standard normally distributed if the fitted model is correct. We test for normality using Jarque-Bera tests.

\subsection*{\small {\it COMPARISON OF GROUP DYNAMIC MODEL TO STANDARD MULTI-STATE RANDOM WALK}}

We also explore the utility of our model in comparison to a standard multi-state random walk. This is done to determine if the additional layer of complexity regarding the centroid movement, and the response of the individuals to the centroid, leads to a more realistic description of observed movement patterns. In particular, we fit a simple 2-state HMM to the reindeer data, with each state involving a correlated random walk. Thus, in both states of the simple HMM, individuals move in complete disregard of the centroid. Parameters were again assumed to be common across individuals.

For each of the 11 individuals and each of the fitted models (standard HMM and group dynamic model), we then simulate 50 movement paths, each starting at the location where the corresponding individual was initially observed, and covering a time period of 200 days. For the group dynamic model we simulate individual-level movement paths using an artificial series of centroid locations that start at the initial (approximate) centroid location, with subsequent movement straight northwards at the constant speed of $0.5$ km/h. This is an essentially arbitrary choice, though it does illustrate the potential usefulness of the suggested approach regarding the modelling of migratory movement.

\section*{Results}

\subsection*{\small {\it SIMULATION EXPERIMENTS}}

All tables giving details on the simulation results are presented in Appendix S1.
In the case where the exact locations of the centroid are known, there is no indication of bias of the parameter estimators. This holds true for both the centroid- and the individual-level (hidden Markov) movement models, and in both Scenarios A and B (cf.\ Tables S1 and S2). Fitting the suggested type of movement model thus is feasible if the centroid's locations are known. %, which in practice may occur for example if groups are led by a particular individual that can easily be identified. 

In the case where the centroid locations can only be approximated, our fitting approach correctly identified the general pattern of the movement at both the individual and the group level, but for both centroid approximation methods the parameter estimates exhibited some bias (cf.\ Tables S1 and S2). For the parameter estimates associated with the individual-level model the bias was generally lower than for the parameter estimates associated with the group-level model. In both cases, the bias was highest (in relative measure) for the concentration parameter of the directional distribution in the BRW component. At the individual level, the concentration parameter corresponds to the individual's bearing relative to the centroid, and this bias is not surprising, since the directional distribution relative to the centroid's location is blurred when using an approximation of the centroid's location. The estimates of the (individual-level) state transition probabilities were also found to be slightly biased, with the fitted models typically predicting more state switches than the true model that was used to simulate the data.  

The results further show that when using approximate centroid's locations, the estimation seems to perform better in cases where the state-dependent distributions are fairly distinct (Scenario A) than in cases where they are not that easily distinguishable (Scenario B). In particular, the accuracy of the Viterbi-based state decoding was higher in Scenario A ($96.1\%$ of the states correctly decoded, vs.\ $95.2\%$ in Scenario B, for the case where the robust mean was applied) (Table S3). Use of the robust mean led to higher accuracy of the state decoding than use of the simple mean ($96.1\%$ vs.\ $94.7\%$ in Scenario A, and $95.2\%$ vs.\ $93.0\%$ in Scenario B). 

\subsection*{\small {\it REINDEER DATA}}

\begin{figure}[!t]
\begin{center}
\includegraphics[width=1\textwidth]{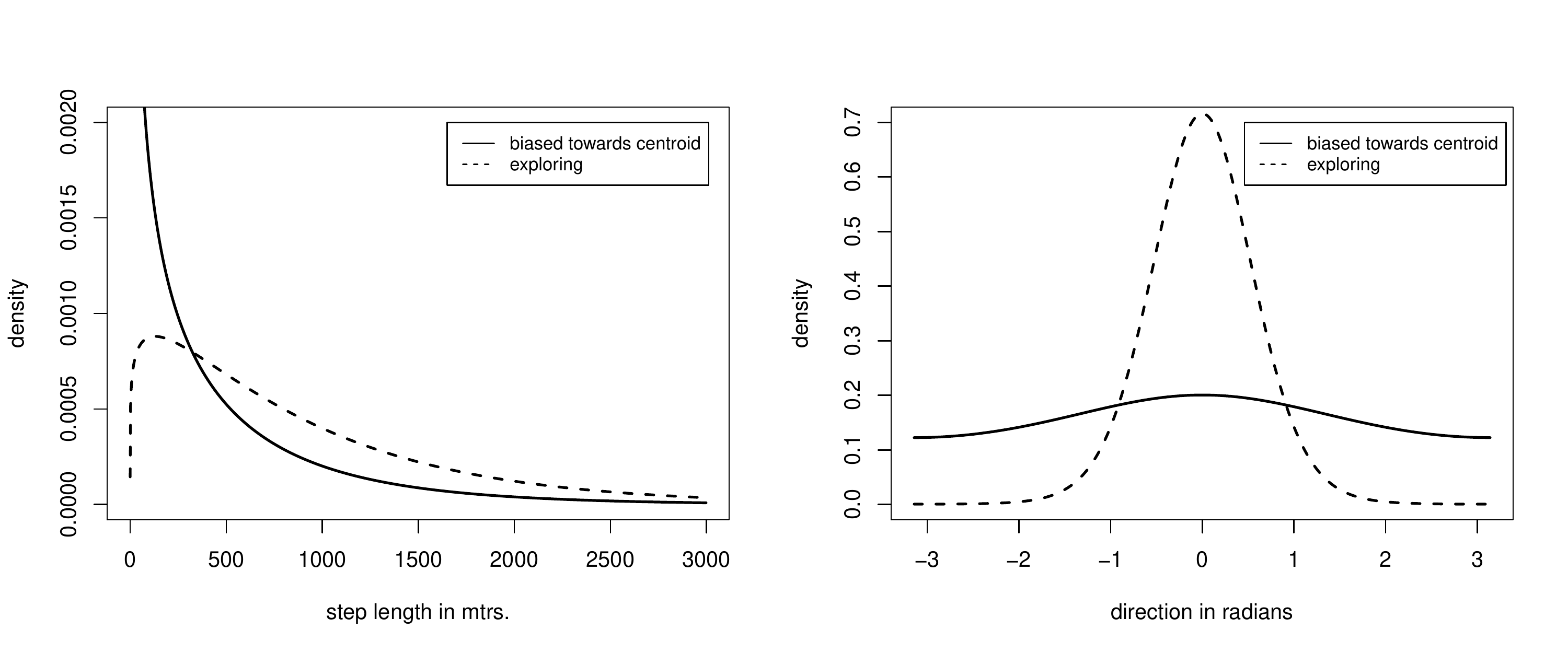}
\end{center}
\vspace{-1.5em}
\caption{Reindeer example, group dynamic model --- fitted state-dependent step length and directional distributions. For the state involving a bias towards the centroid (state 1), radian 0 corresponds to the direction the centroid is in, relative to the individual. For the ``exploring'' state (state 2), radian 0 corresponds to the previous movement direction of the individual, i.e., while exploring the individuals tend to move straight ahead. \label{steps}}
\end{figure}

Parameter estimates for the group dynamic model fitted to the reindeer data, together with 95\% CIs obtained based on the Hessian of the log-likelihood, are listed in Table \ref{estims}. The fitted state-dependent distributions for step lengths and directions are displayed in Figure \ref{steps}. In 22 Jarque-Bera tests for normality of the pseudo-residuals, the null hypothesis of normality is twice rejected at the 5\% level, which is consistent with the model fitting well.

\begin{table}[!htb]
\centering
\caption[Caption]{Parameter estimates for the group dynamic individual-level model, fitted to the reindeer data, and 95\% confidence intervals of the estimates (obtained based on the Hessian of the log-likelihood).}\label{estims}\vspace{0.5em}
% \vspace{1em}
\begin{tabular}{l l c c c}
\hline\\[-0.9em]
Parameter       &  Description & CI lower bound         & estimate & CI upper bound  \\
\hline\\[-0.5em]
state 1: & & & \\ 
$\mu_1$         & {\it gamma mean} & 399.5 & 432.9 & 469.1 \\
$\sigma_1$      & {\it gamma standard deviation} & 524.8 & 571.2 & 621.8 \\
$\kappa_1$      & {\it von Mises concentration} & 0.183 & 0.246 & 0.331 \\
$\gamma_{11}$   & {\it persistence in state 1} & 0.830 & 0.866 & 0.895 \\[0.4em]
state 2: & & & \\ 
$\mu_2$         & {\it gamma mean} & 805.0 & 896.6 & 998.6 \\
$\sigma_2$      & {\it gamma standard deviation} & 743.8 & 829.7 & 925.5 \\
$\kappa_2$      & {\it von Mises concentration} & 2.759 & 3.517 & 4.482 \\
$\gamma_{22}$   & {\it persistence in state 2} & 0.510 & 0.584 & 0.655 \\
 \hline\\[-0em]
\end{tabular}
\end{table}  

In state 1, individuals on average make relatively short steps (gamma mean $=432.9$) and are weakly attracted to the group's centroid (von Mises concentration $=0.246$), while in state 2, individuals on average make longer steps (gamma mean $=896.6$) and perform a CRW with high directional persistence (von Mises concentration $=3.517$). Following \citet{mor04}, we interpret state 1 as the behavioural state in which animals are ``encamped'', here associated with a weak attraction to the group centroid, and state 2 as the behavioural state in which animals are ``exploring''. Such interpretations should not be taken literally, as they merely provide a rough classification of different movement patterns (\citealp{lan12}). In particular, encamped does not necessarily mean with the group; rather it means the animal is making shorter steps and those steps are slightly biased towards the group centroid (Figure \ref{steps}). %Nevertheless, it is interesting that the model indicates that the reindeer cover longer distances when moving solitarily than when being attracted to the group centroid. 

\begin{figure}[!t]
\includegraphics[width=1\textwidth]{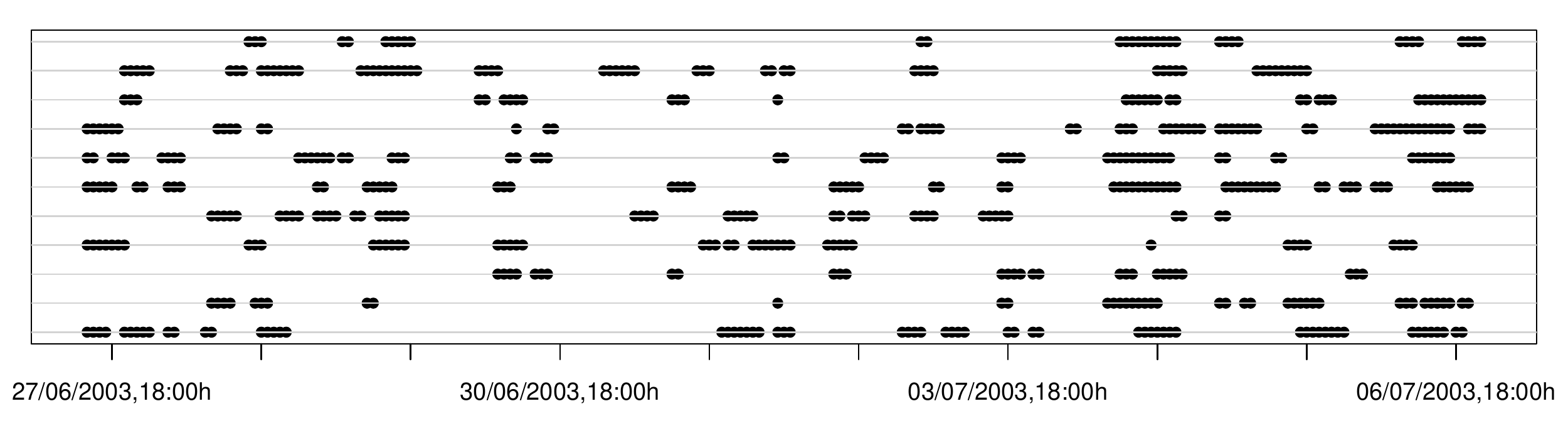}
\vspace{-2em}
\caption{Reindeer example --- Viterbi-decoded state sequences for the 11 different individuals. Each grey horizontal line corresponds to one individual, with black dots indicating when the corresponding individual was estimated to be in state 2 of the fitted model. \label{viterbi}}
\end{figure}

\begin{figure}[!htb]
\begin{center}
\includegraphics[width=0.9\textwidth]{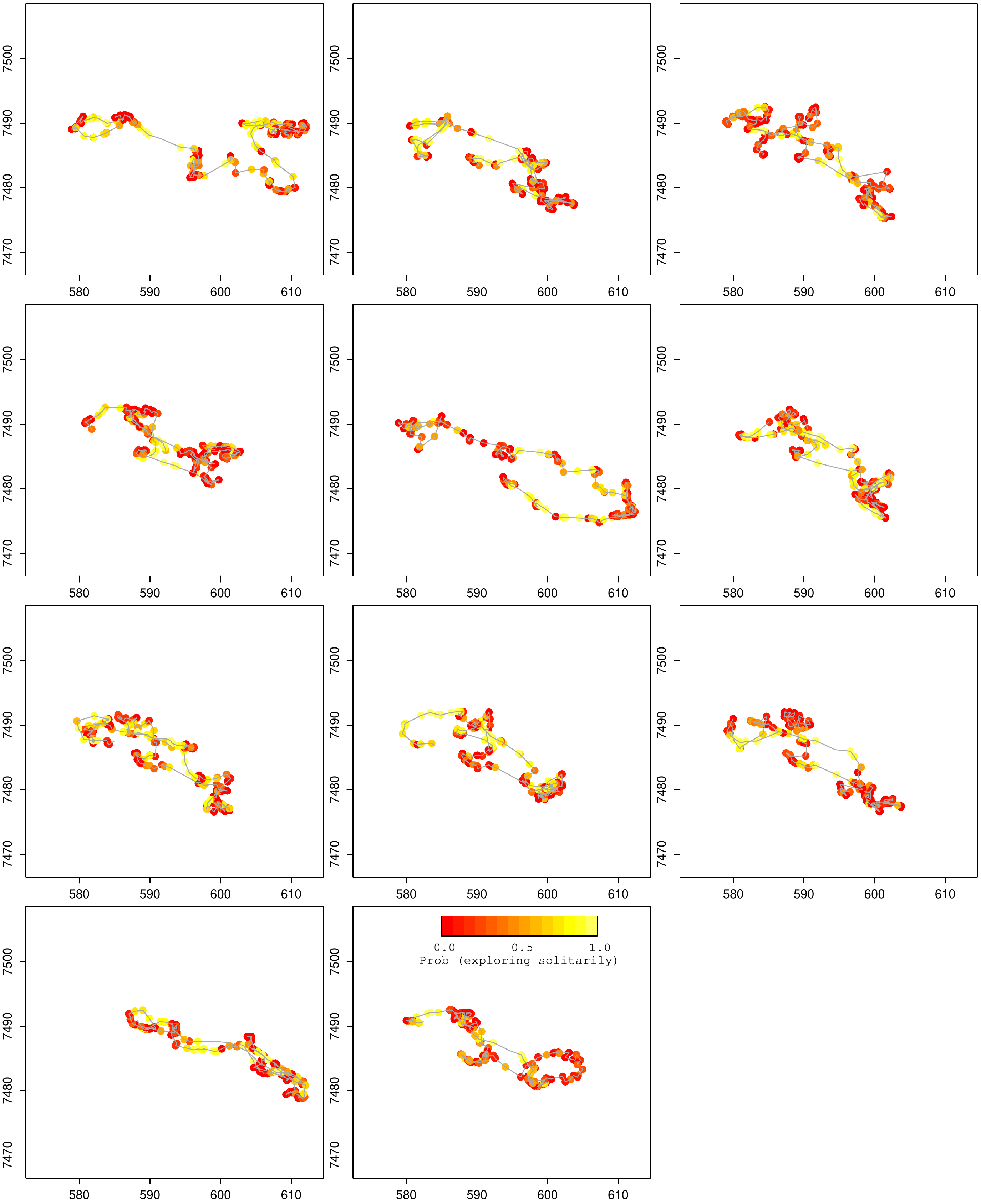}
\end{center}
\vspace{0em}
\caption{Reindeer example --- state categorizations mapped onto the 11 movement paths. The colour of the points mapped on the locations indicate the probability of the corresponding animal being in state 2 (``exploring''), i.e., red indicates higher probability of being encamped and attracted to the group centroid. The horizontal and vertical axes give $x$ and $y$, respectively, in kilometres using the Sweref99 projection from the National Land Survey of Sweden. } 
\label{stateprobs}
\end{figure}

According to the stationary distribution of the fitted Markov chain, approximately three quarters of the time is spent in state 1. Figure \ref{viterbi} displays the Viterbi-decoded state sequences, i.e., the sequences of states that are most likely to have given rise to the observations. Figure \ref{stateprobs} displays the estimated probabilities of being in the exploratory state (and thus moving in disregard of the group centroid), mapped onto the movement paths, for the 11 reindeer. Broadly speaking, the individual reindeer spend long stretches of time encamped (Figure \ref{viterbi}), and these periods are interspersed with brief exploratory forays (shown in yellow in Figure \ref{stateprobs}).

\subsection*{\small {\it GROUP DYNAMIC MODEL VS.\ STANDARD MULTI-STATE RANDOM WALK}}

\begin{figure}[!htb]
\begin{center}
\includegraphics[width=0.8\textwidth]{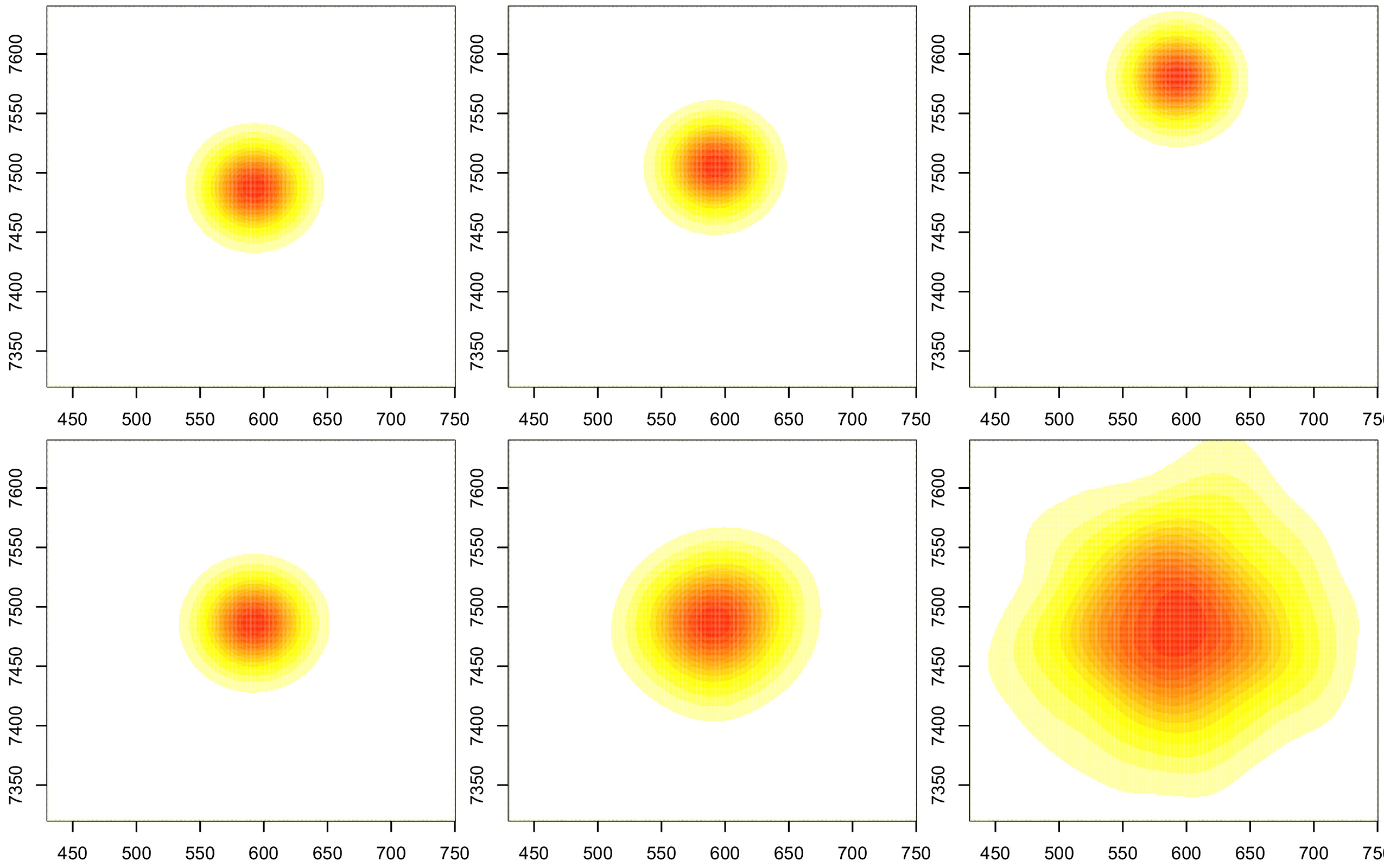}
\end{center}
\vspace{-0em}
\caption{Reindeer example --- spatial densities of simulated locations of individuals after $9.4$ days (left column), 50 days (middle column) and 200 days (right column), for the group dynamic model (top row; with artificial centroid movement straight northwards) and for the standard two-component mixture of correlated random walks (bottom row). The smoothed densities were obtained by applying a bivariate kernel density estimator to the generated locations. Within a short time span (left column) the two models provide similar predictions both in terms of mean and variance of the densities. For simulation over longer time periods (middle and right column), the attraction mechanism in the group dynamic model (top row) both bounds the extreme movements of the individuals (variance) and ensures that individuals follow the group centroid (mean), while the two-state model without attraction (bottom row) is purely diffusive. \label{spatial}}
\end{figure}

The centroid-driven individual-level model yields a much lower value for the Akaike Information Criterion (AIC) than the simple two-state random walk ($\Delta$AIC$=43.6$). However, the AIC has to be interpreted very cautiously here: the additional centroid layer appears only in the former model, and the stated $\Delta$AIC is based on regarding the centroid's positions as deterministic covariates within the individual-level model. In addition to the statistical preference (decreased AIC) for the centroid-driven model, we also illustrate the difference in predictive capacity between the two models using simulation (Figure \ref{spatial}). Although this is a simplified example, it highlights the effect of the group attraction mechanism, which ensures that individuals stick together and follow the group (Figure \ref{spatial}, top row) rather than wander independently and randomly about (Figure \ref{spatial}, bottom row), as is the case for the simple two-state CRW. 

\section*{Discussion}\label{discuss}

\subsection*{\small {\it GENERAL REMARKS ABOUT THE APPROACH}}

Through simulation and by fitting our model to real data, we have documented that reliable inference can be made on the influence of the group on the movement dynamics of social individuals. The suggested approach provides a tractable way to understand how the role of social factors, such as dominance hierarchies and gregariousness, influence the movement decisions of a group, and adds to a growing body of literature \citep{cou11,dow11,pol10,pol11}. The key advancement in this approach that distinguishes it from previous work is its inferential nature.  The likelihood-based framework lends itself to including environmental or physiological parameters (such as forage conditions, season, or reproductive status) as potential explanatory covariates in the fission-fusion dynamics of communally living organisms, which makes itself immediately applicable to ecological research. Unlike commonly done in the SPP literature on the movement of social groups \citep{cou05}, here we fit our model to actual movement data. Furthermore, we are able to statistically compare the fit of competing models (model selection) and to assess the goodness of fit of our model (model checking). %We illustrated this by fitting the group dynamic model to movement data from reindeer and found that it provided a superior description of the data as opposed to a two-state random walk model without individual interaction. We checked the goodness of fit of the group dynamic model using pseudo-residuals, and found no evidence for the model being inadequate. While this was a small illustrative example, our results do indicate the inferential potential of the presented framework for group dynamic movement.

The key component of our framework that enables it to capture group behaviour is the attraction to the centroid. Specifically this is done by allowing individuals to switch between a ``group'' state, where individuals are biased toward the group centre, and one or more ``exploratory'' states where individual movement is independent of the group. In contrast to multi-state CRWs, the fitted group dynamic movement model is non-diffusive. More precisely, even when the attraction to the centroid is weak, individuals will, in the long run, not move arbitrarily far away neither from the centroid nor from each other (cf.\ Figure \ref{spatial}).

One pre-existing approach that is similar in spirit to ours is that of \citet{dun77}, who formulate all of their models in terms of multiple interacting animals. They think of the position of $m$ animals in $d$ dimensions (taking $d=2$ in practice, as we do here) as a single point in $k=m\times d$ dimensions, and model the movement of that point using a multivariate Ornstein-Uhlenbeck process, parameterised in terms of a $k$-vector and two $k\times k$ matrices. By contrast, our approach can be thought of as modelling the position of the group in terms of a point and having individuals interact directly with the centroid. Their models have very large numbers of parameters already for moderate values of $m$, and in practice our representation is far more parsimonious and interpretable, certainly for $m>2$.

Hierarchical models are an alternative way to capture similarities between individuals, by modelling the model parameters as stochastic variables with a common group or population mean \citep{jon06}. Thus, individuals' parameters are correlated, but conditional on those parameters there is no interaction between individuals. We assumed here that all individuals shared movement parameters, and focused on the interaction of their movement paths via the group centroid. In contrast to the traditional hierarchical approach, this models a fundamental mechanism of the herding behaviour and therefore has a more intuitive biological interpretation. 

%It is not clear what types of data are required for reliable inference in different scenarios. 
Using multiple yet relatively short real data time series, we were able to reveal an attraction to the centroid. However, it is difficult to make general statements about how many individuals need to be tracked, or for how long, to make realiable inference on attraction to the centroid. It is also unclear as to how influential the proportion of tracked animals to untracked animals needs to be in terms of reliably estimating the centroid position, but it is intuitive that the model will work best for relatively cohesive groups, where the choice of particular individuals does not bias the analysis. The feasibility of the group centroid concept clearly also depends on the complexity of the fission-fusion dynamics; e.g., the concept of a single group centroid will sometimes be inadequate.

\subsection*{\small {\it DISCUSSION OF RESULTS}}

We generated and analysed simulated data to assess the ability of our modelling approach to consistently estimate model parameters and to distinguish group behaviour from solitary behaviour in individuals' movement. When the true location of the group centroid was known, the approach provided unbiased parameter estimates of both the centroid movement model and the individual-level movement model. In cases where the movements of an alpha individual direct the movement of the group, the true centroid location is known when this animal is identified and tagged. This is harder for other species where individual roles in the group are less clear prior to tagging. For scenarios in which the centroid location is unknown, we considered simple or robust means of the individuals' locations as approximations to the centroid's locations. In our simulations, the estimation of the individual-level movement models worked reasonably well when the centroid's locations were approximated using the simple mean, and slightly better when using the robust mean. In addition, we were able to use the centroid attraction information to estimate behavioural states with high accuracy. Our model thus represents a useful approach to statistical and ecological inference on group movement dynamics.

We illustrated the modelling approach using movement data from 11 reindeer in Sweden \citep{ska08,ska10}. We were able to infer, to some extent, when individuals are tracking group movement, and when they appear to be following their own movement drivers (cf.\ Figure \ref{steps}). %In general, it appears there are several spatial areas where the group seems cohesive (e.g. $(x,y)=(600,7480)$ and $(x,y)=(590,7485)$; cf.\ Figure \ref{stateprobs}). 
The models we fitted to the reindeer data led to the classification that is characteristic of two-state random walks, with an encamped and an exploratory state and corresponding state-dependent step lengths and turning angle distributions \citep{mor04,lan12}, except that in the group dynamic model the encamped state involves the additional feature of a weak attraction to the group centroid. According to the fitted model and based on Viterbi-decoding of the underlying state sequences, some individuals spent periods of up to three days in the nominal encamped state during the time period covered by the observations. 

Taken together with the simulation results, fitting our model to real data highlights the ecological inference that can be gleaned from our approach. First, we have quantified how the individual reindeer respond to the group. This highlights how conspecific attraction can be included in movement models. Second, we have shown how behavioural states can be estimated with high accuracy, even in cases where important movement phenomena, i.e., step lengths, are similar across states. Finally, we have outlined advantages of including the centroid information over simple multi-state random walks. Specifically, including the centroid layer in the model has led to a more accurate depiction of spatial density,  highlighting the importance of incorporating social drivers.  

\subsection*{\small {\it EXTENSIONS}}

In the simulation experiments, estimators were biased when we had to approximate the centroid's locations. As an alternative, the centroid could be approximated using the movement directions of the individuals, rather than their locations. This approximation is closer to the data-generating system and might therefore have reduced bias, but it requires knowledge about which individuals exhibit group behaviour at a given time. Another alternative could be to use a suitable function of the individuals' locations and movement as a noisy proxy for the location of the centroid. Combining this with a movement model, the location of the centroid could be estimated within a state-space modelling framework, e.g., using a Bayesian data augmentation approach for the centroid location. Such an approach is appealing since the uncertainty about the centroid's location would be propagated through to the individual-level parameter estimates, thus potentially providing more accurate parameter estimates and confidence bounds. 

A simplifying assumption of our method was that all individuals shared movement parameters. A more realistic model could include random effects for certain parameters such as the switching probabilities, thereby allowing for variations in behaviour between individuals \citep{for12,sch12}, i.e., that some individuals are more adventurous than others. Random effects, however, describe individual differences as stochasticity (or noise) and hence do not explain the source of the variation. Instead, formulating individual-level parameters as functions of auxiliary information such as, e.g., gender, weight or height, would enable testing of hypotheses related to individual covariates and aid in uncovering reasons for differences between individuals within the group.

Similarly, the model can easily be extended to include internal and external dynamic covariates. For example, nonhomogeneous Markov chains could be considered for the behavioural state dynamics, with transition probabilities structured such that observations of the ambient environment or body condition mediate switches between movement phases. Such a model may enable separation of the relative roles of group membership, individual body condition and habitat, and represents an exciting avenue of analysis for the movement patterns of social animals. An important special case of such nonhomogeneous Markov chains results from including time as a covariate, which can be useful in order to accommodate seasonality or the daily cycle in the model. Another potentially interesting covariate to consider is the individual-specific separation distance from the centroid, which could be informative about group membership. From the statistical point of view, these extension are fairly straightforward, since the likelihood structure of the HMM remains unaffected.

Another direction of development is to extend the continuous-time models of \citet{dun77} to incorporate some of the advantages of the current approach. Explicitly representing the centroid as part of the process means that the movement of the whole system is then a diffusion process in $(m+1)\times d$ dimensions, but the modelling of that process is greatly simplified if the $m$ individual animals are each interacting with the centroid, as here, rather than interacting directly with each other. One possibility currently being pursued is for each animal to follow a multivariate Ornstein-Uhlenbeck (MOU) process attracted to the centroid, while the centroid follows its own movement model. Under simplifying assumptions, the whole system can then be modelled as a partially observed $(m+1)\times d$-dimensional MOU process.

\section*{Acknowledgements}
Funding for RL was provided by the Engineering and Physical Sciences Research Council (ESPRC reference EP/F069766/1). Funding for RSS was provided by US Office of Naval Research grant N00014-12-1-0286 to University of St Andrews. Funding for PGB and MN was provided by EPSRC/NERC grant EP/1000917/1. The ideas presented in this manuscript were developed during a workshop on animal movement modelling, held at the University of St Andrews in June 2012, which was funded via an EPSRC Strategic Partnership Fund. %The authors would like to thank Tiago Marques, a referee and an associate editor for very helpful comments on previous versions of the manuscript. 

\makeatletter
\renewcommand\@biblabel[1]{}

\markboth{}{}
\bibliographystyle{plainnat}

\end{linenumbers}
\end{spacing}

 \newpage

\section*{Appendix S1} 

\setlength{\parindent}{0pt}

\renewcommand{\thetable}{S\arabic{table}}
\setcounter{table}{0}

\begin{table}[!htb]
\centering
\caption[Caption]{Simulation results -- means and standard deviations (in brackets) of parameter estimates for centroid movement model, as obtained in $500$ simulation runs, for centroid approximation methods TL (true location), SM (simple mean) and RM (robust mean). The considered movement model is a (single-state) BCRW, with step length mean parameter $\mu^{(c)}$, step length standard deviation parameter $\sigma^{(c)}$, directional distribution concentration parameter $\kappa^{(c)}$, and $\eta^{(c)}$ the weight of the CRW component in the BCRW (such that $1-\eta^{(c)}$ is the weight of the BRW component; see Langrock {\em et al.}, 2012).}\label{simtab1}\vspace{1em}
\begin{tabular}{l r r r r}
\hline\\[-0.9em]
                &  true value & TL         & SM    & RM  \\
\hline\\[-0.5em]
{\bf Scenario A} & & & \\[0.1em] 
$\mu^{(c)}$         & 15   & 14.99 (0.486)  & 13.87 (0.408) &  17.39 (0.657)  \\
$\sigma^{(c)}$      & 10   & 9.96  (0.476)  & 7.27 (0.282)  &  9.48 (0.732) \\
$\kappa^{(c)}$      & 1    & 1.01  (0.088)  &  0.65 (0.099) &   0.27 (0.079) \\
$\eta^{(c)}$        & 0.85 & 0.855 (0.082)  & 0.905 (0.090) &  0.822 (0.164) \\[0.8em]
{\bf Scenario B} & & & \\[0.1em] 
$\mu^{(c)}$         & 15   & 15.01 (0.491)  & 13.75 (0.378) &  17.10 (0.452)  \\
$\sigma^{(c)}$      & 10   & 10.03  (0.458) &  7.13 (0.292) &   9.13 (0.462) \\
$\kappa^{(c)}$      & 1    & 1.01  (0.085)  &  0.62 (0.091) &   0.26 (0.077) \\
$\eta^{(c)}$        & 0.85 & 0.854 (0.083)  & 0.895 (0.097) &  0.809 (0.169) \\[0.5em] \hline\\[-0.9em]
\end{tabular}
\end{table}

\begin{table}[!htb]
\centering
\caption[Caption]{Simulation results -- means and standard deviations (in brackets) of parameter estimates for individual-level movement model, as obtained in $500$ simulation runs, for centroid approximation methods TL (true location), SM (simple mean) and RM (robust mean). The considered movement model is a two-state HMM (see main manuscript for details).}\label{simtab2}\vspace{1em}
\begin{tabular}{l c c c c}
\hline\\[-0.9em]
                &  true value & TL         & SM    & RM  \\
\hline\\[-0.5em]
{\bf Scenario A} & & & \\[0.4em] 
state 1: & & & \\ 
$\mu_1$         & 30   & 30.00 (0.224) & 30.01 (0.224) &  29.98 (0.224)  \\
$\sigma_1$      & 15   & 15.01 (0.211) & 15.02 (0.211) &  15.00 (0.212) \\
$\kappa_1$      & 2.5  & 2.50 (0.051)  & 1.15 (0.180)  &  2.10 (0.064) \\
$\gamma_{11}$   & 0.99 & 0.990 (0.002) & 0.988 (0.003) &  0.988 (0.002)  \\[0.4em]
state 2: & & & \\ 
$\mu_2$         & 50   & 49.94 (1.572) & 48.21 (2.474) &  48.95 (1.760)  \\
$\sigma_2$      & 25   & 24.91 (1.291) & 24.93 (1.324) &  24.90 (1.319) \\
$\kappa_2$      & 5    & 5.02  (0.465) & 5.03 (0.558)  &  4.98  (0.496) \\
$\gamma_{22}$   & 0.85 & 0.846 (0.028) & 0.829 (0.040) &  0.824 (0.038)  \\[0.8em]
{\bf Scenario B} & & & \\[0.4em] 
state 1: & & & \\ 
$\mu_1$         & 30   & 30.00 (0.221) & 30.11 (0.247) &  30.03 (0.222)  \\
$\sigma_1$      & 15   & 15.02 (0.207) & 15.06 (0.220) &  15.02 (0.208) \\
$\kappa_1$      & 2.5  & 2.50 (0.051)  & 1.50 (0.186)  &  2.12 (0.061) \\
$\gamma_{11}$   & 0.99 & 0.990 (0.002) & 0.982 (0.006) &  0.985 (0.003)  \\[0.4em]
state 2: & & & \\ 
$\mu_2$         & 30   & 29.96 (0.978) & 30.11 (1.051) &  30.03 (0.938)  \\
$\sigma_2$      & 15   & 14.93 (0.817) & 14.46 (0.836) &  14.83 (0.796) \\
$\kappa_2$      & 5    & 5.04  (0.471) & 4.28 (0.833)  &  4.78 (0.536) \\
$\gamma_{22}$   & 0.85 & 0.846 (0.029) & 0.791 (0.048) &  0.803 (0.046)  \\[0.5em] \hline\\[2.9em]
\end{tabular}
\end{table}

\begin{table}[!htb]
\centering
\caption[Caption]{Simulation results -- mean proportions of the individual's states that were correctly decoded by the Viterbi algorithm, which was applied in each of $500$ simulation runs, based on models fitted using the centroid approximation methods TL (true location), SM (simple mean) and RM (robust mean), respectively.}\label{simtab3}\vspace{1em}
\begin{tabular}{l c c c c}
\hline\\[-0.9em]
                & TL         & SM    & RM  \\
\hline\\[-0.5em]
{\bf Scenario A} & 99.4\% & 98.8\% & 99.1\% \\[0.4em] 
{\bf Scenario B} & 99.2\% & 97.6\% & 98.7\% \\[0.5em] 
\hline\\[-0.9em]
\end{tabular}
\end{table}

\end{document}